\documentclass[final,1p,times]{elsarticle}
\biboptions{numbers,sort&compress}
\usepackage{amsmath}
\usepackage{xcolor}
\usepackage{float}
\usepackage{subcaption}
\usepackage{amssymb}
\usepackage{amsthm}
\usepackage{multirow}
\usepackage[bookmarks,bookmarksnumbered]{hyperref}
\hypersetup{colorlinks = true,linkcolor = blue,anchorcolor =red,citecolor = blue,filecolor = red,urlcolor = red,pdfauthor=author}
\usepackage{hyperref,graphicx}       
\begin{document}

\begin{frontmatter}


 \title{A new iterative algorithm for generating gradient directions to
detect white matter fibers in brain from MRI data}
 \author{Ashishi Puri\corref{cor1}\fnref{label1}}
 \ead{apuri@ma.iitr.ac.in}
 \author{Sanjeev Kumar\fnref{label1}}
 \ead{sanjeev.kumar@ma.iitr.ac.in}
\cortext[cor1]{Corresponding author}
 \address[label1]{Department  of  Mathematics,   Indian  Institute  of  Technology  Roorkee,   Roorkee,247667, India}
 



\begin{abstract}
This paper proposes an iterative algorithm for choosing gradient directions use to reconstruct
white matter fibers in the brain. The present study is not focusing on data acquisition where
scanning is performed. The Adaptive Gradient Directions (AGD) [1] approach is extended to
refine the position and area of the grid, resulting in an admissible reduction in angular error.
We begin with the gradient directions distributed uniformly inside a grid of bigger size and with
larger spacing between the points. Both (size of the grid and spacing between the points) reduce
iteratively. The proposed algorithm ensures that the actual position of fiber comes inside the
grid at each iteration, unlike as in the AGD approach. As a result, the solution tends to actual
orientation in each iteration followed by better estimation of fibers. The proposed algorithm is
validated by associating it with mixture of Gaussian diffusion and mixture of non-central Wishart
distribution models. The proposed approach significantly reduce the angular error for multiple
computer-generated experiments on synthetic simulations and real data. Moreover, we have also
performed simulations with fibers not residing in the XY-plane. For this set-up also, the proposed
work outperforms, giving lesser angular error with both the models. Synthetic simulations have
been performed with Rician distributed (R-D) noise of standard deviation ($\sigma$) ranging from $\sigma =
0.02-0.1$. This work helps in better understanding of the anatomy of the brain using the MRI
signal data.

\end{abstract}


\end{frontmatter}

\section{\textbf{Introduction}}

Magnetic Resonance Imaging (MRI) is a non-invasive and in-vivo neuroimaging technique that provides the anatomical arrangement of tissue microstructure of the diverse organs. The random Brownian motion of water molecules inside tissue voxels leads to MR signal loss. Diffusion weighted MRI is typically based on this Brownian motion of water molecules. Basser et al. in 1994 \cite{basser1994mr} introduced the paradigm of Diffusion Tensor Imaging (DTI) as a revolution in the medical field. The brain contains billions of neurons forming a neural network for communicating with each other \cite{mori2006principles}. DTI uses the anisotropic diffusion of water molecules to reconstruct or replicate the organization of fibers in the nervous system. The fundamental ideas, implementations, and principles related to DTI have already been discussed in several articles earlier \cite{mori1999diffusion, luypaert2001diffusion, mori2006principles, jones2011diffusion}. Isotropic diffusion is observed in case of gray matter fibers and cerebrospinal fluid as signal is independent of the direction in which the gradients are applied. However, water diffusion is anisotropic in white matter fibers as diffusion is directionally dependent for water diffuses unrestricted along axon but encounters restriction perpendicular to the direction of axon \cite{soares2013hitchhiker}.\\

DTI is capable of calculating the orientation and rate of the diffusion in a voxel of tissue possessing a single fiber but malfunctions in the case where multiple fibers are crossing each other in one voxel \cite{alexander2007diffusion}. Various techniques for acquisition of image and reconstruction paradigm have been introduced that are progressively reducing the error \cite{westin2014measurement}. Probabilistic mixture models viz. mixture of Gaussian distribution (MoG) \cite{tuch2002high}, mixture of central Wishart distribution (MoCW) \cite{jian2007multi, jian2007novel} and mixture of non-central Wishart distribution (MoNCW)  \cite{shakya2017multi, shakya2017multitrac} are skilled to reconstruct crossing fibers, unlike DTI. These models provide a great platform for knowing the thorough and clear anatomy of the brain. Recent work on achieving high anatomical accuracy of reconstructed images have been done in \cite{schilling2020brain, schucht2020visualization}\\

Among these models, the MoNCW performed extensively well in distinguishing the fibers. In \cite{jian2007multi,shakya2017multi}, uniformly distributed fixed number of points (gradient directions) over a unit sphere have been considered for reconstruction of fibers. Later on, a new promising technique for choosing the gradient directions is introduced by Puri et al. in \cite{puri2021enhanced} called the Adaptive Gradient Directions (AGD). This technique neither fixes the value for the number of gradient directions, nor the directions are uniformly distributed over the surface of unit sphere in contrary to what has been considered in \cite{jian2007multi,shakya2017multi}. AGD algorithm when compared with uniform gradient directions (UGD) approach, the former shows huge reduction in angular error while distinguishing the mixture of fibers' orientation in a voxel. However, the dimension of the grid \cite{puri2021enhanced} as well as the spacing between adaptive gradient directions play a vital role in obtaining the actual orientation of fibers. The limitation of increasing or decreasing the area of the grid or spacing between the points have already been discussed in \cite{puri2021enhanced}. When following the optimal case of the AGD approach where gradient directions per grid are fixed to 49, there is quite high probability of losing the actual orientation of fiber from entering inside the grid. This may lead to a disoriented depiction of the direction of fibers.\\

In the present study, we aim to investigate the upper bounds
of reconstruction performance to achieve a high anatomical accuracy of reconstructed brain image. We introduce a novel technique for generating gradient directions for reconstructing or replicating the white matter fibers' orientation in brain. We have come up with new iterative approach for depicting the fiber orientation as an extension of what has been done in AGD. A larger value of $K$ gives lesser angular error and vice-versa. But choosing large $K$ results in multiple directions that are redundant for the region of interest leading to high angular error, computationally expansive etc \cite{puri2021enhanced}. Here we begin with uniformly distributed $K=20$ points over a unit sphere in step-1. The result that we obtained in step-1 has a significantly less accuracy rate as the value as $K$ is very small. Then considering the rough orientation that we have obtained in step-1, we form a grid of large area containing uniformly distributed points with larger spacing. In step-2, we work on these points (directions) to find the result approaching the actual orientation.
In the succeeding steps or iterations, we again form a new grid centered at the point obtained in step-2 but with a smaller area and  a higher density of points. Following this way, leads to the actual fiber orientation that we need to detect. The proposed iterative algorithm is capable of reducing the acquired angular error procured using the AGD approach. The proposed algorithm has been coupled with MoG and MoNCW model. Associating the proposed iterative approach for gradient directions with these models enhances the efficiency and rate of accuracy of the models to great extent. Our

\section{\textbf{Methods}}
\subsection{Preliminaries}

\subsubsection{DTI Model}

DTI model based on Gaussian (Normal) diffusion consists of Stejskal-Tanner (S-T) equation \cite{stejskal1965spin} was capable of estimating fibers in voxels containing single fiber. However, this model is not practiced to detect fibers in the regions of orientational heterogeneity. The `b-value' also called diffusion weighting $b$ is given as $b=(\gamma\delta\textbf{G})^2t$, where $\gamma$ is the gyromagnetic ratio, $\delta$ is diffusion gradient time and $t$ is the effective diffusion time. Also $G$ and $\textbf{g}$ denote magnitude and direction of the diffusion sensitizing gradient $\textbf{G}$ respectively. Displacement probability assumed by DTI is specified by Gaussian probability distribution function. The well-established ST-equation is given by,

\begin{equation} \label{eq:Eq. 1}
S(\textbf{q}) = S_0\exp(-b\textbf{g}^T\mathbb{D}\textbf{g})
\end{equation}

where $S_0$ is the signal intensity in the absence of any weighting gradient, $\mathbb{D}$ is $3\times 3$ diffusion matrix, and $\textbf{q}$ is the coordinate vector in q-space given as $\textbf{q}=\frac{1}{2\pi}\gamma\delta\textbf{G}$. DTI model was not practised for voxels with crossing fibers, this restriction was eliminated when mixture models viz. Mixture of Gaussian diffusion (MoG) \cite{tuch2002high}, MoCW and MoNCW were introduced. For a detailed description of  these models interested reader is referred to \cite{tuch2002high, jian2007multi, shakya2017multi}.

\subsubsection{MoG Model} 

For resolving the orientational heterogeneity in the brain voxels, MoG model  \cite{shakya2017multi} is introduced where each voxel is split into $K$ multiple compartments such that $w_i$ is the weight (or volume) fraction of $i^{th}$ compartment. Signal intensity equation based on Gaussian diffusion for multi-fiber case is given by

\begin{equation} \label{eq:Eq. 2}
\frac{S(\textbf{q})}{S_0} = \sum_{i=1}^{K} w_i \exp(-b\textbf{g}^T\mathbb{D}_i\textbf{g})
\end{equation}

 where, $K$ is the mixing components or the compartments of the voxel and $\mathbb{D}_i$ is the diffusion tensor corresponding to the $i^{th}$ mixing component of the voxel.
 
\subsubsection{MoNCW Model }

Assuming that the diffusion in brains' voxels is following a non central Wishart distribution, MoNCW model is introduced \cite{shakya2017multi}. To establish the MoNCW model, an additional non-centrality parameter denoted by $\Omega \in Sym_+^d $ was introduced in the MoCW model, where $Sym_+^d$ denotes the manifold of symmetric positive definite matrices. Assuming $\textbf{Y}_{n\times p}=[Y_1,Y_2,...,Y_n]^T$ be a $n\times p$ matrix. Here all the rows are independently taken from a $p$-variate normal distribution, $Y_i\sim N(\mu_i, \Sigma)$. Calculating a matrix $\textbf{P}_{n\times p}$ using the population means $\mu_i$ such that $\textbf{P}=[\mu_1,\mu_2,...,\mu_n]^T$. Then the random symmetric matrix defined by $\textbf{F}=\textbf{Y}^T\textbf{Y}$ has a non-central Wishart (NCW) distribution , $W_p(n,\small{\Sigma},\Omega)$. Here the expectation of random matrix $\textbf{F}$ is  E$(\textbf{F})=n\Sigma+\Omega$ \cite{james1955non, li2003noncentral, letac2004tutorial, pham2015trace}.\\
 
Considering $\Omega = 0$, the above reduces to the central Wishart distribution, $W_p(n,\Sigma)$. Further using the Laplace transform of NCW distribution \cite{mayerhofer2013existence}, the signal intensity equation based on NCW distribution is given as follows \cite{shakya2017multi}:

\begin{equation} \label{eq:Eq. 3}
S(\textbf{q})/S_0= \sum_{i=1}^K w_i(1+trace(\Sigma_i\textbf{B}))^{-n}\exp~[-trace(\textbf{B}\{I_p+\Sigma_i\textbf{B}\}^{-1}\Omega_i)]
\end{equation}

where, $n\geq0$ is the shape parameter, $\Sigma\in sym_+^d$ is the scale parameter, $\Omega$ is the non-centrality parameter and $\textbf{B}=b\textbf{gg}^T$. Using expected value of NCW distribution $(\textbf{D} = n\Sigma+\Omega)$, we have $\mathbb{D}_i=n\Sigma_i + \Omega_i$ \cite{shakya2017multi}. The parameter $\Omega_i$ is calculated using this relation. Shakya et.al. \cite{shakya2017multi} used an ad-hoc approach  to estimate the best value for $\alpha$. As a result, $\alpha=0.99$ was chosen for further calculations.\\

Eq. \ref{eq:Eq. 2} and Eq. \ref{eq:Eq. 3} reduces  to following linear system of equations:

\begin{equation} \label{eq:Eq. 4}
A\textbf{w} = \textbf{Y}+\eta
\end{equation}

where $\textbf{Y}=S(\textbf{q})/S_0$ is signal vector that has been normalized, $\eta$ is the noise and every entry of matrix $A$ is given as,\\

$$
    A_{ji}= 
\begin{cases}
   exp(-b\textbf{g}_j^T\mathbb{D}_i\textbf{g}_j),& \textrm{for MoG model}\\
    (1   + trace(\Sigma_{i}{\textbf{B}}_j))^{-n}\exp~[-trace\{\textbf{B}_j(I_p + \Sigma_i \textbf{B}_j)^{-1}
\Omega_i\}],              & \textrm{for MoNCW model}
\end{cases}
$$

where, $j=1,2,...,N$ and $N$ denotes the number of gradient directions along which signals have been generated. The vector $\textbf{w} = \{(w_i): i=1,2,...,K\}$ in Eq. \ref{eq:Eq. 4}, is the unknown parameter that has to be calculated. Here, $K$ denotes the mixture components or number of gradient directions used in reconstruction process. $w_i$ is the mixture weights corresponding to the $i^{th}$ compartment of the voxel.  We used the Non-Negative Least Square
(NNLS) method \cite{lawson1995solving} to solve the linear system of equations given in Eq. \ref{eq:Eq. 4} in the proposed work.\\

In \cite{jian2007multi, shakya2017multi}, uniformly distributed $K=321$  gradient directions have been used to estimate the fibers. As mentioned earlier, using such large value for $K$ is computationally rigorous  and time consuming approach. This also leads to extremely large number of redundant directions resulting in multiple unnecessary calculations. As a result, AGD approach is introduced in \cite{puri2021enhanced} which is successful in reducing error along with computation time. Moreover, it also eliminated the unnecessary calculations by working only on those directions that have the maximum probability for existence of the fibers. \\

\subsubsection{Adaptive Gradient directions}

\begin{figure}[!ht]
  \centering
 {\includegraphics[width=3in]{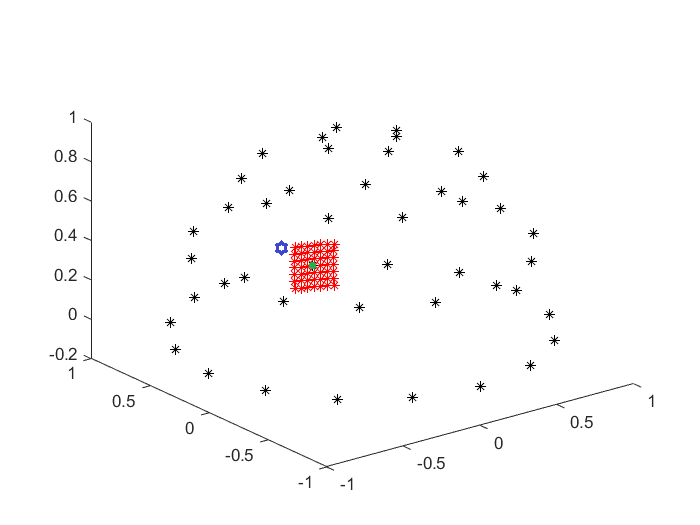}}
  \caption{Gradient directions used in AGD approach with blue colored star as the actual position of fiber.}
  \label{fig:1}
\end{figure}

AGD approach  \cite{puri2021enhanced} consists of two steps. In the first step, uniformly distributed 46 gradient directions are used in place of $K$ in Eq. \ref{eq:Eq. 4}. Then obtaining the rough idea about the fiber orientation in this step, we generated AGD in the next step. These AGD are uniformly distributed in the neighbourhood of the orientations achieved in the first step. Then using these AGD vectors in place of $K$ in Eq. \ref{eq:Eq. 4} give the results for AGD approach. Although this algorithm, when compared with UGD (uniformly distributed points over a surface of unit sphere), gave lesser angular error but we encounter one limitation in this procedure.
 The restriction on the size of grid, as explained in \cite{puri2021enhanced}, resulted in the exclusion of the actual position of the fiber from the grid. This leads to comparatively higher inaccuracy in reconstructing the fibers. 
In Fig \ref{fig:1}, the blue colored star represents the actual fiber orientation and the green colored dot indicates the rough fiber orientation that we obtained in the first step of AGD approach. Black and red colored stars depict the gradient directions used in step-1 and AGD, respectively. 

\subsection{Proposed iterative algorithm for generating gradient directions}

The `$K$' vectors along the gradient directions are the uniformly distributed points on unit sphere used in reconstruction process in the state-of-art models. The value of $K$ is directly proportional to the rate of accuracy in finding the orientation of fibers. However, using larger $K$ is computationally rigorous and time consuming since the algorithm has to be executed along $K$ directions. Also AGD approach may result in a situation where the actual orientation lies outside of the grid, as shown in Fig \ref{fig:1}. In this figure, if blue colored star is the actual position of fiber, then due to restriction in the size of the grid, it is excluded from the grid. This will then result in disoriented estimation of fibers followed by a high angular error. To resolve this drawback, we develop an iterative approach to obtained the solution closest to the actual fiber orientation. The proposed iterative algorithm is discussed below: \\

\textbf{Step-1:} We solve the Eq. \ref{eq:Eq. 4} for uniformly distributed $K$ components over the surface of unit hemisphere to obtain pairs of azimuthal ($\phi$) and polar $(\theta)$ angles. We use a very small value for $K=20$ that can provide us an idea about the rough position of fibers.\\

\begin{figure}[!ht]
  \centering
  \subcaptionbox{\label{fig2:a}}{\includegraphics[width=1.8in]{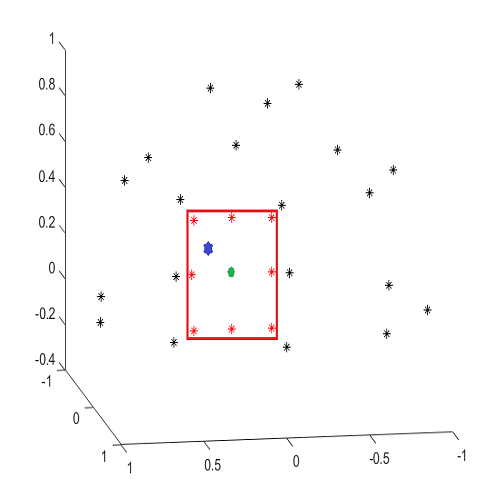}}
  \subcaptionbox{\label{fig2:b}}{\includegraphics[width=1.7in]{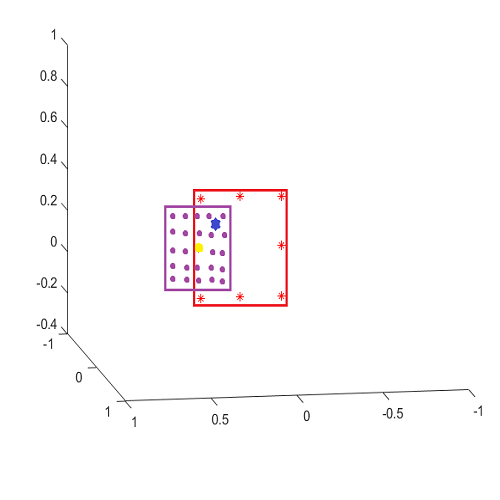}}
  \subcaptionbox{\label{fig2:c}}{\includegraphics[width=1.7in]{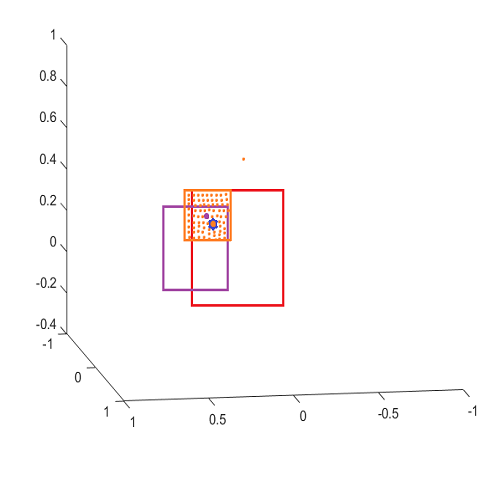}}
  \caption{Gradient directions used in the iterative algorithm of proposed approach with different boundary colors representing grids formed at each stage }
  \label{fig:2}
\end{figure}

For each such pair, we have spherical coordinates $(\phi_{k_1},\theta_{k_1},r)$ with fixed radial distance $r=1$ and  $k_1 \in \mathbb{Z}^{+}$ denoting the number of rough fiber orientations obtained in Step-1. In Fig. \ref{fig2:a}, the green colored dot represents this point where $k_1=1$ is considered for simplicity.\\

\textbf{Step-2:} Now, we generate the uniformly distributed gradient directions/points in the neighbourhood of the $k_1$ points obtained in step-1. These points are located over the hemisphere forming a grid centered at each $k_1^{th}$ orientation. For this, new pairs of azimuthal and polar angles are generated using the following steps:\\

\hspace{0.5cm} \textbf{Step-2.1:} Set of newly developed azimuthal angles = $\{\phi_{k_1},~\phi_{k_1} \pm m_1, ~\phi_{k_1} \pm 2m_1, .~.~.~, \phi_{k_1} \pm c_1m_1\}$
such $c_1, m_1 \in \mathbb{Z}^+$ where $c_1$ accounts for the number of points inside the grid and $m_1$ accounts for the spacing between the points.\\

\hspace{0.5cm} \textbf{Step-2.2:} Set of newly developed polars angles = $\{\theta_{k_1},~\theta_{k_1} \pm m_1, ~\theta_{k_1} \pm 2m_1, .~.~.~, \theta_{k_1} \pm c_1m_1\}$.\\
 
\hspace{0.5cm} \textbf{Step-2.3:} Set of new pairs of azimuthal and polar angles   = $\bigg\{(\phi_{k_1},\theta_{k_1}), (\phi_{k_1}, \theta_{k_1} \pm m_1),~.~.~.~,~$

$(\phi_{k_1},\theta_{k_1}\pm c_1m_1),~(\phi_{k_1}\pm m_1,\theta_{k_1}),~(\phi_{k_1}\pm m_1,\theta_{k_1} \pm m_1),~ .~ .~ .~ ,~(\phi_{k_1}\pm m_1,\theta_{k_1} \pm c_1m_1),~ .~ .~ .~ ,~$

$(\phi_{k_1}\pm c_1m_1,\theta_{k_1}),~(\phi_{k_1}\pm c_1m_1,\theta_{k_1} \pm m_1), ~. ~. ~.~, ~ ~(\phi_{k_1}\pm c_1m_1,\theta_{k_1} \pm c_1m_1)\bigg\} $\\

The total number of directions obtained in this step per grid = $(2c_1+1)^2$\\
The total number of directions obtained in this iteration i.e.  $K'$ = $(2c_1+1)^2\times k_1$\\

Now we convert these $K'$ spherical coordinates with radial distance unity to Cartesian coordinate system and these vectors will then replace the $K$ mixing components required in Eq. \ref{eq:Eq. 4}. These points are represented by the red colored stars in Fig. \ref{fig2:a}. \\

\textbf{Step-3:} Solving the linear system of equations for above generated  $K'$ vectors in Eq. \ref{eq:Eq. 4}, we obtained $k_2$ number of orientations. The $k_2$ pairs of azimuthal and polar angle we obtained are now be treated as centres for the grids that we will generated in the next step. This centre orientation is represented by yellow dot in the purple colored grid in Fig. \ref{fig2:b}. For simplicity, we have considered $k_2=1$.\\

\textbf{Step-4:}  The $k_2$ number of grids are formed each centred as mentioned earlier. The new pairs of azimuthal and polar angles are calculated as follows: \\

\hspace{0.5cm} \textbf{Step-4.1:} Set of newly generated azimuthal angles = $\{\phi_{k_2},~\phi_{k_2} \pm m_2, ~\phi_{k_2} \pm 2m_2, .~.~.~, \phi_{k_2} \pm c_2m_2\}$
such $c_2, m_2 \in \mathbb{Z}^+$ where $m_2<m_1$ and $c_2	> c_1$.\\

\hspace{0.5cm} \textbf{Step-4.2:} Set of newly generated polars angles = $\{\theta_{k_2},~\theta_{k_2} \pm m_2, ~\theta_{k_2} \pm 2m_2, .~.~.~, \theta_{k_2} \pm c_2m_2\}$.\\
 
\hspace{0.5cm} \textbf{Step-4.3:} Set of new pairs of azimuthal and polar angles   =  $\bigg\{(\phi_{k_2},\theta_{k_2}), (\phi_{k_2}, \theta_{k_2} \pm m_2),~.~.~.~,~$

$(\phi_{k_2},\theta_{k_2}\pm c_2m_2),~(\phi_{k_2}\pm m_2,\theta_{k_2}),~(\phi_{k_2}\pm m_2,\theta_{k_2} \pm m_2),~ .~ .~ .~ ,~(\phi_{k_2}\pm m_2,\theta_{k_2} \pm c_2m_2),~ .~ .~ .~ ,~$

$(\phi_{k_2}\pm c_2m_2,\theta_{k_2}),~(\phi_{k_2}\pm c_2m_2,\theta_{k_2} \pm m_2), ~. ~. ~.~, ~ (\phi_{k_2}\pm c_2m_2,\theta_{k_2} \pm c_2m_2)\bigg\} $\\

The total number of directions obtained in this step per grid = $(2c_2+1)^2$\\
The total number of directions obtained in this iteration i.e.  $K''$ = $(2c_2+1)^2\times k_2$. \\

Now replacing $K'$ with  $K''$  in Eq. \ref{eq:Eq. 4} lead us to the orientation of fibers reaching closer to the actual fiber orientation.
The $K''$ points are represented by the purple coloured dots inside the purple grid in Fig. \ref{fig2:b}. \\

 Repeating the same process leads us to the results that have been reported in this paper. The fiber's actual position is represented by the blue colored star in Fig. \ref{fig2:a}, which is not one of the positions that we chose in step-1. Using a bigger size of the grid in the first iteration makes sure that the actual position comes inside the grid, unlike as in AGD algorithm where due to restrictions in the grid size this position remains out of the grid  resulting in comparatively higher angular error. In Fig. \ref{fig2:b}, we observe that again the actual position is entering the grid formed in the next iteration. Moreover, this position is coming closer and closer to the vectors that we choose to work upon, ensuring reduction in angular error at each iteration. In Fig. \ref{fig2:c}, the purple colored star inside the orange grid  indicates the position of fiber acquired in the previous step and orange colored dots represent the generated gradient directions in a form of grid (centred at purple dot) neighbouring the purple dot uniformly. The overlapping of the blue star with the orange colored dot in Fig. \ref{fig2:c} validate the proposed algorithm. More results on synthetic simulations and real data have been discussed in this paper in the next section. \\
 
\section{\textbf{Results}}

\subsection{\textbf{Parameter setting and Error metric}}

All the synthetic simulation experiments have been performed using $46$ gradient directions i.e. $N=46$ to generate the signal intensity $S(\textbf{q})$. Following Shakya et al. \cite{shakya2017multi}, we set $n=2$ ($n$ belongs to Gindikin ensemble $\Lambda=\{\frac{a}{2}:a=1,2,...,(p-2)\}\cup\big[\frac{p-1}{2},\infty))$ \cite{gindikin1975invariant, n1988davidson, peddada1991proof}. We have used $b=1500s/mm^2$ and for UGD approach, we used $K=321$. In order to obtain the weight vector $\textbf{w} = \{(w_i): i=1,2,...,K\}$ in all the required equations, we used non-negative least square method. Synthetic data is generated using MATLAB$^{TM}$ open library \cite{barmpoutis2009adaptive, barmpoutis2010tutorial}. Here adaptive kernels were considered for approximating exact continuous diffusion weighted -MRI signal.\\

We have defined resultant angular error for every simulations as the sum of error in azimuthal and polar angles. For all the simulations, polar angle is assigned a fixed value of $90^\circ$. To calculate angular error (A.E.) we use the below mentioned formula:

$$A.E. = \frac{\sum_{k=1}^{I}|A_k^{original}- A_k^{estimated}|}{I}$$

where, $I$ denotes total number of fibers per voxel and $A$ is replaced by $\phi$ and $\theta$ for calculating their respective angular errors. Computer generated experiments have been conducted with R-D noise. Considering the standard deviation of noise denoted by $\sigma$, noise is added to the signal intensity vector $Y$ of synthetic simulations as follows:

$$\textbf{Y}=\sqrt{(\textbf{Y}+\sigma*randn(1))^2+(\sigma*randn(1))^2}$$
\vspace{0.1cm}

where `randn' function helps in generating  random numbers that are distributed normally. In this article, simulations have been primarily performed with R-D noise having standard deviation, $\sigma=0.06$. Table~\ref{table:1} and \ref{table:2} represent the resultant angular error procured with different separation angles between $2$ and $3$ crossing fibers respectively. The errors in bold represent the minimum error obtained using 4 different approaches. For all the cases (except for $70^\circ$), the calculated error is minimum when proposed we followed the proposed algorithm. \\

\begin{table}[!ht] 
\parbox{.45\linewidth}{
\centering
    \begin{tabular}{|c||c|c|c|c|}
    
   \hline
  $|\phi_1-\phi_2|$ & MoNCW with & MoNCW with & MoG with  & MoG with \\
  (in degrees) & AGD &  proposed algorithm & UGD & proposed algorithm\\
  
\hline

40 & 10.17 & \textbf{8.7} & 13.79 & 11\\

\hline

50 & 9.25 &\textbf{ 7.5} & 9.71 & \textbf{7.5}\\

\hline

60 & 6.9 & \textbf{3.1} & 4.36 & 3.43 \\

\hline

70 & 3.4 & 2.5 & \textbf{2} & 3.3\\

\hline

80 & 2.1 & \textbf{1.1}& 2.5 &2\\

\hline

90 & 1 & \textbf{0} & 1.5 & 0.5\\

\hline

100 & 0.9 & \textbf{0.5 }& 2.07 & 0.71\\

\hline

110 & 2.8 & 2 & 1.86 & \textbf{1.71}\\

\hline

120 & 3.67 & \textbf{2.4} & 3.64 & 2.64\\

 \hline\hline
\end{tabular}

}
 \caption{Comparison of resultant angular error of $\phi$ and $\theta$ using different appraoches with R-D noise, $\sigma=0.06$ for 2-crossing fibers in a voxel. For all simulations polar angle is set to $90^{\circ}$}\label{table:1}
  
\end{table}

\begin{table}[!ht] 
\parbox{.45\linewidth}{
\centering
    \begin{tabular}{|c||c|c|c|c|}
    
   \hline
  $|\phi_1-\phi_2|$& MoNCW with & MoNCW with & MoG with  & MoG with \\
  (in degrees) & AGD &  proposed algorithm & UGD & proposed algorithm\\

\hline

 (10,70,130) &  1.5 & \textbf{0.17}&2.52&0.81\\

\hline

(0,40,80) & 17.37  & 14.4&16.14&\textbf{12}\\

\hline

(10,60,110) &   7.3  & \textbf{5.6}&8.43&6.09\\

\hline

(8,67,118) & 8.2 & 3.5&3.52&\textbf{2.81}\\

\hline

(15,55,110) &  13.1 & \textbf{9.3}&15&12.41\\

\hline

(50,110,170) &  2.23 & 1.3&5&\textbf{0.23}\\

\hline

(10,50,90) &  14.17 & 14.37&\textbf{12.05}&13.43\\
 
\hline

 (40,105,160) & 5.67 & 1.07&3.83&\textbf{0.90}\\
 
 \hline

 (0,65,130) & 5.62 & \textbf{2.66} &4.35&3.36\\
  \hline\hline
\end{tabular}
}
 \caption{Comparison of resultant angular error of $\phi$ and $\theta$ using different appraoches with R-D noise, $\sigma=0.06$  3-crossing fibers  in a voxel. For all simulations polar angle is set to $90^{\circ}$}\label{table:2}
  
\end{table}

\subsection{\textbf{Visual results for synthetic simulations on fiber separation}}

\begin{figure}[!ht]
  \centering
  \subcaptionbox{\label{fig3:a}}{\includegraphics[width=2.09in]{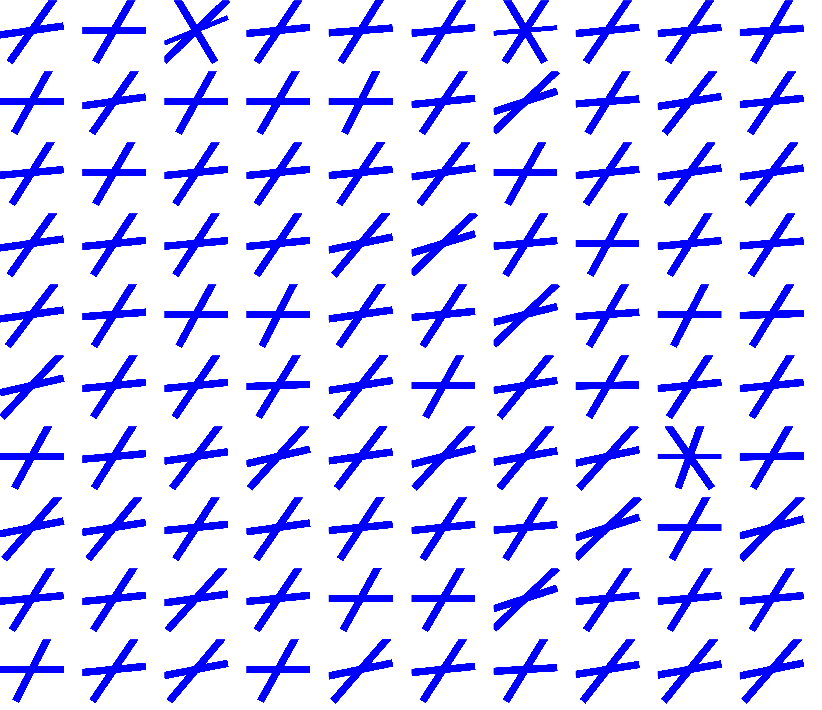}}\hfill
  \subcaptionbox{\label{fig3:b}}{\includegraphics[width=2.09in]{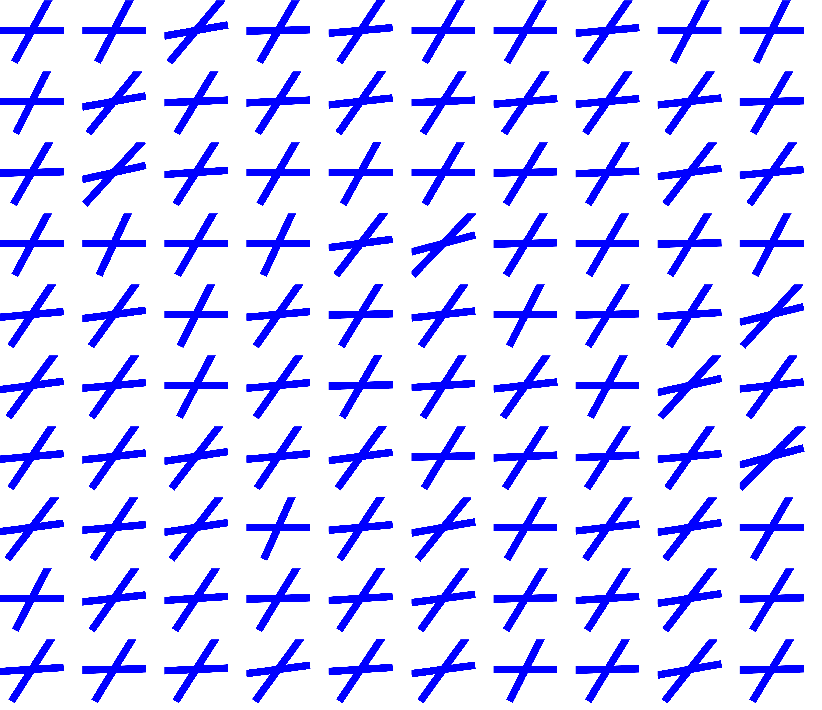}}
   \caption{Estimation of fibers using (a) MoNCW with AGD and (b) MoNCW with proposed technique for 2-crossing fibers with $(\phi_1,\phi_2)=(0^\circ,60^\circ)$ and noise, $\sigma=0.1$. The weight fraction corresponding to both the orientations is set to 0.5.}
  \label{fig:3}
\end{figure}

\begin{figure}[!ht]
  \centering
  \subcaptionbox{\label{fig4:a}}{\includegraphics[width=2.3in]{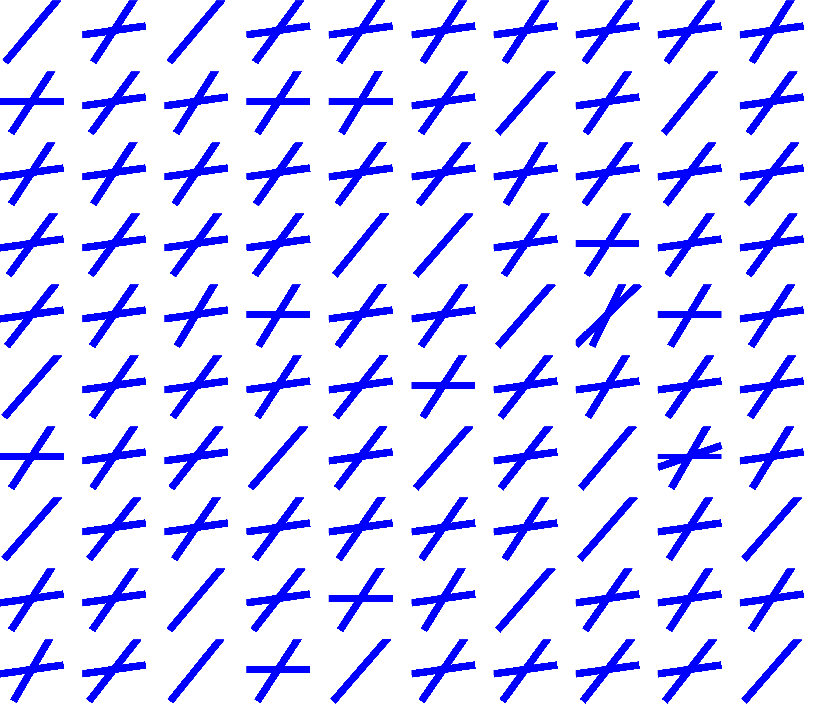}}\hfill
  \subcaptionbox{\label{fig4:b}}{\includegraphics[width=2.3in]{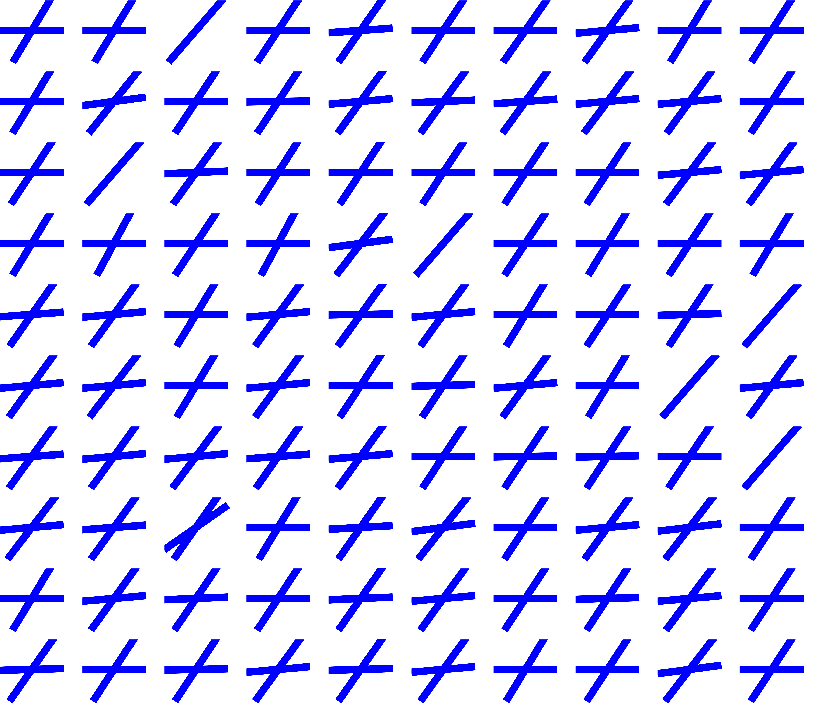}}
    \caption{Estimation of fibers using (a) MoNCW with AGD and (b) MoNCW with proposed technique for 2-crossing fibers with $(\phi_1,\phi_2)=(0^\circ,60^\circ)$ and noise, $\sigma=0.06$. The weight fraction corresponding to these fibers is set to 0.3 and 0.7 respectively.}
  \label{fig:4}
\end{figure}

We begin with 100 simulations of 2 crossing fibers oriented at $(\phi_1,\phi_2)=(0^\circ,60^\circ)$ with R-D noise, $\sigma=0.1$. In addition, in this simulation we considered equal weight fraction for both
the orientations i.e. 0.50 vs. 0.50 respectively. The performance of AGD and proposed algorithm is shown in Fig. \ref{fig:3}. Next we carried out 100 simulations of the previous orientations with noise, $\sigma=0.06$ and with unequal weight fraction i.e. 0.30 vs. 0.70 for $0^\circ$ and $60^\circ$ respectively. The results have been shown in Fig. \ref{fig:4}. As shown in figure, $0^\circ$ orientation having lesser weight value, is not reconstructed in many of the simulations using AGD approach unlike the proposed approach. Following the same trend, we performed 100 simulations with same noise such that fibers are oriented at $0^\circ$ and $60^\circ$ having weights $0.2$ and $0.8$ respectively.  \\

After these experiments consisting of 2 orientations in two directions, we carried out simulations on 3 crossing fibers such that $(\phi_1,\phi_2, \phi_3)=(0^\circ,65^\circ,135^\circ)$. The visual results have been displayed in Fig. \ref{fig:6}. Highly noisy data have been generated with R-D noise of $\sigma=0.1$ and weight fractions for all the orientations is set equal i.e. $(\frac{1}{3})$. Since the separation angle between the fibers is quite large, so both the approaches are capable of estimating all the three fibers but disorientation of $0^\circ$ can be clearly observed in AGD approach. Next, we carried out the comparison between MoG with UGD approach and with the proposed approach as shown in  Fig. \ref{fig:7}. We performed $100$ simulations for 3-crossing fibers oriented at $(\phi_1,\phi_2, \phi_3)=(15^\circ,55^\circ,110^\circ)$. Here also proposed algorithm works well in separating the fibers efficiently in most of the simulations when compared with MoG models with UGD approach. \\

\begin{figure}[!ht]
  \centering
  \subcaptionbox{\label{fig5:a}}{\includegraphics[width=2.3in]{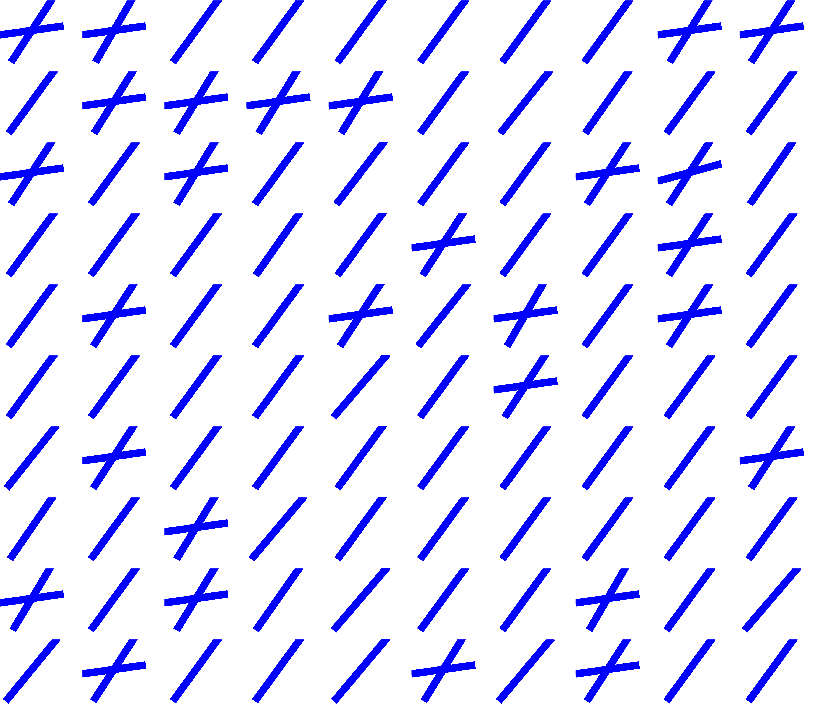}}\hfill
  \subcaptionbox{\label{fig5:b}}{\includegraphics[width=2.3in]{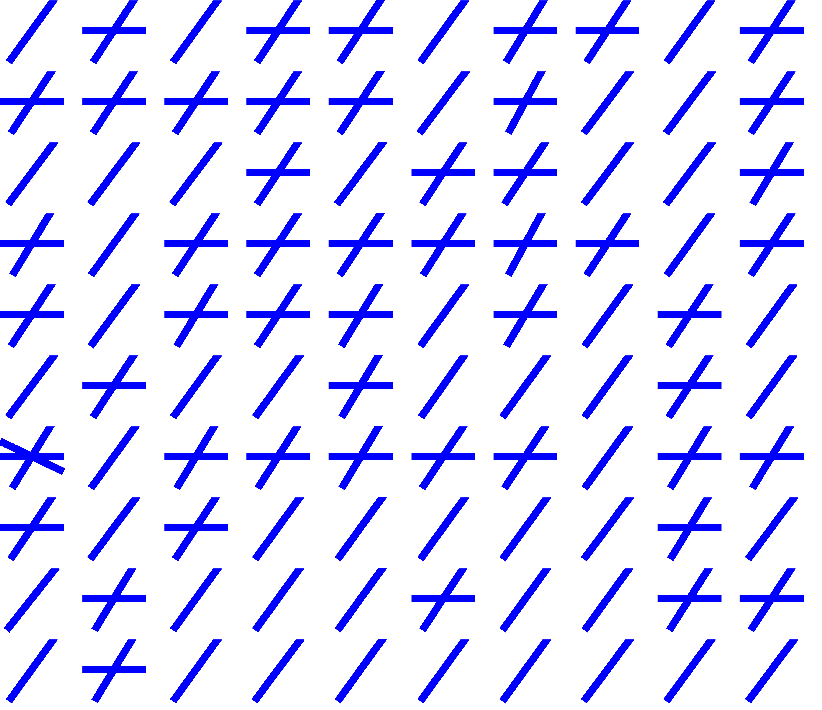}}
    \caption{Estimation of fibers using (a) MoNCW with AGD and (b) MoNCW with proposed technique for 2-crossing fibers with $(\phi_1,\phi_2)=(0^\circ,60^\circ)$ and noise, $\sigma=0.06$. The weight fraction corresponding to these fibers is set to 0.2 and 0.8 respectively.}
  \label{fig:5}
\end{figure}

\begin{figure}[!ht]
  \centering
  \subcaptionbox{\label{fig6:a}}{\includegraphics[width=2.5in]{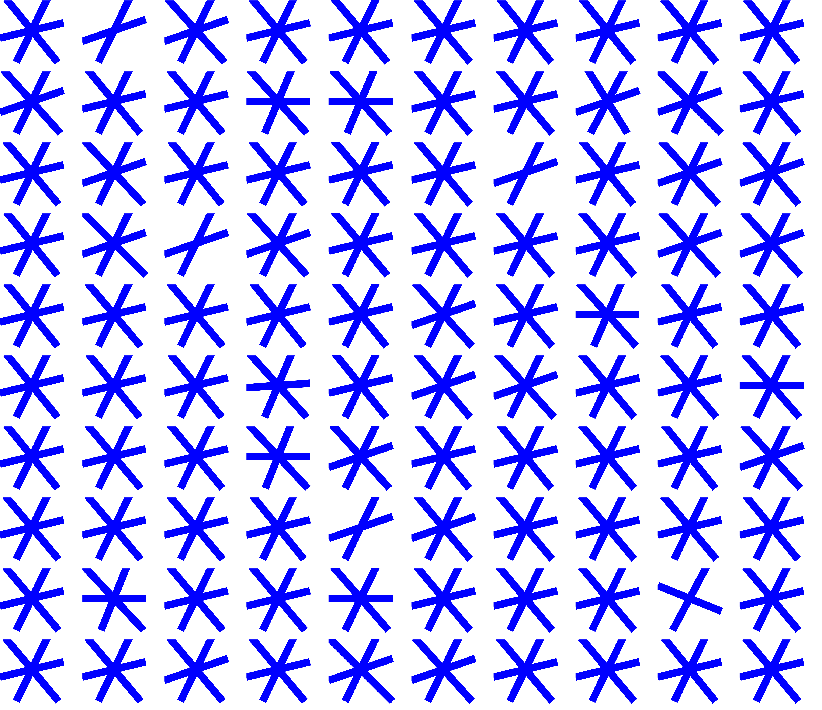}}\hfill
  \subcaptionbox{\label{fig6:b}}{\includegraphics[width=2.5in]{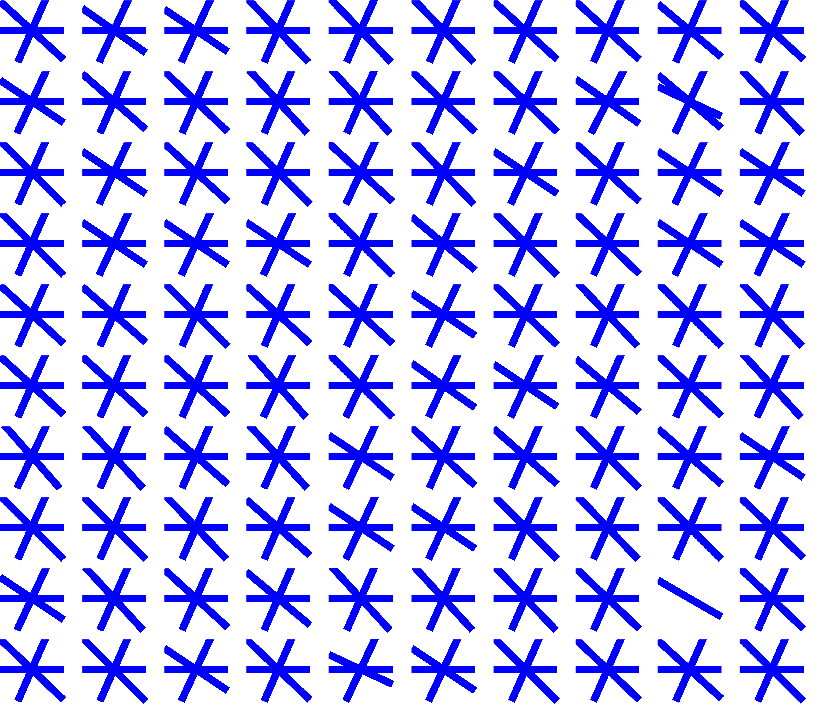}}
    \caption{Estimation of fibers using (a) MoNCW with AGD and (b) MoNCW with proposed technique for 3-crossing fibers with $(\phi_1,\phi_2,\phi_3)=(0^\circ,65^\circ,130^\circ)$ and noise, $\sigma=0.1$. The weight fraction corresponding to each fiber is set to $\frac{1}{3}$.}
  \label{fig:6}
\end{figure}

\begin{figure}[!ht]
  \centering
  
  \subcaptionbox{\label{fig7:a}}{\includegraphics[width=2.5in]{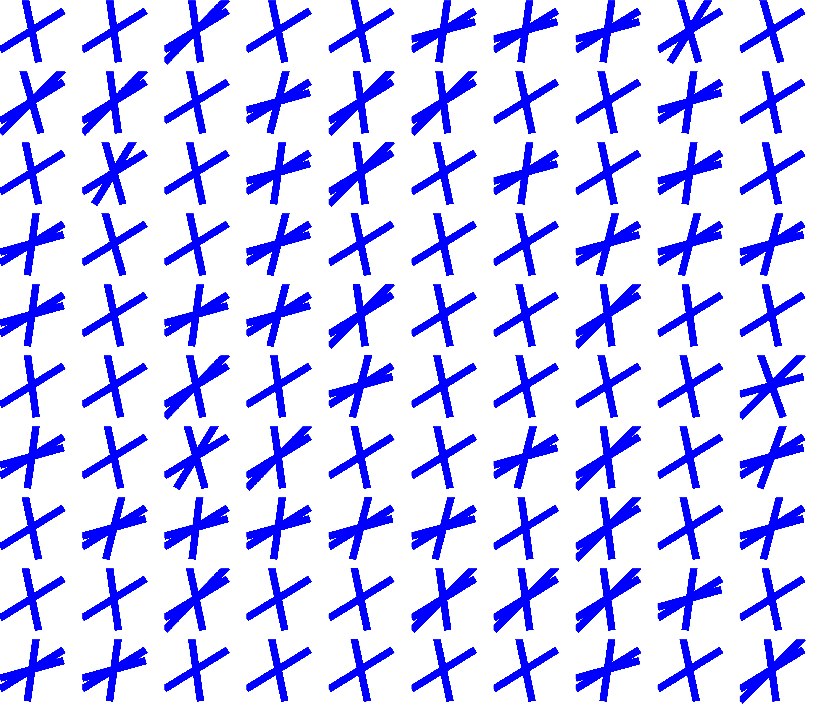}}\hfill
  \subcaptionbox{\label{fig7:b}}{\includegraphics[width=2.5in]{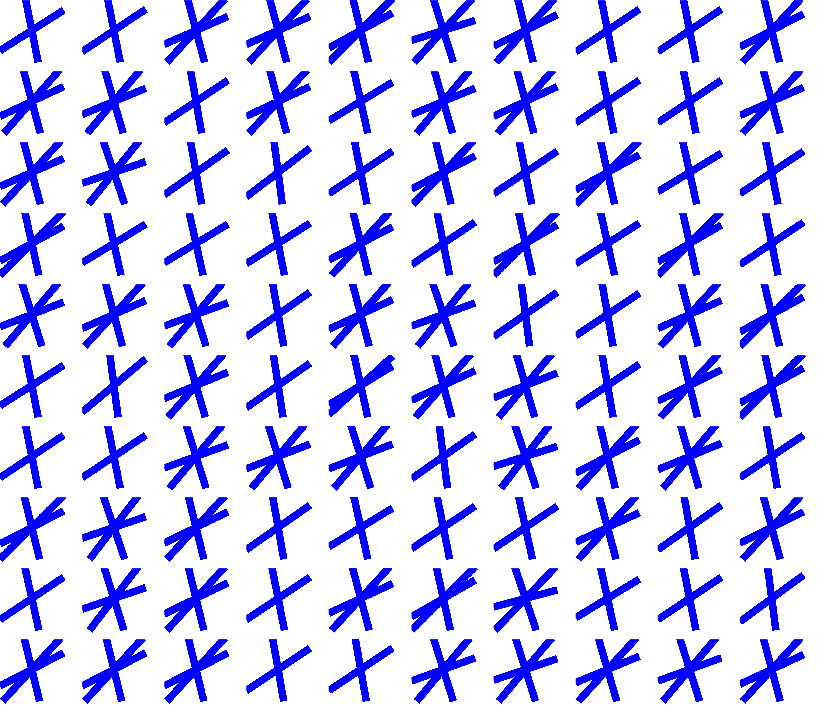}}
  \caption{Estimation of fibers using (a) MoG with UGD and (b) MoG with proposed technique for 3-crossing fibers with $(\phi_1,\phi_2,\phi_3)=(15^\circ,55^\circ,110^\circ)$ and noise, $\sigma=0.06$. The weight fraction corresponding to each fiber is set to $\frac{1}{3}$.}
  \label{fig:7}
\end{figure}

\begin{figure}[!ht]
  \centering
  
  \subcaptionbox{\label{fig8:a}}{\includegraphics[width=2.1in]{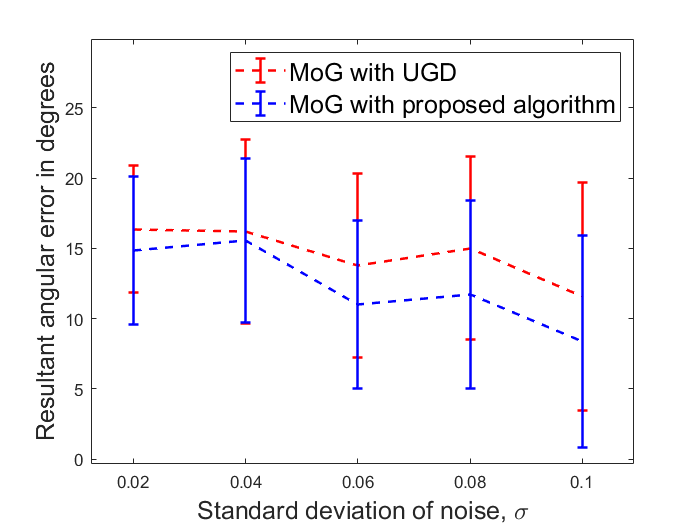}}
  \subcaptionbox{\label{fig8:b}}{\includegraphics[width=2.1in]{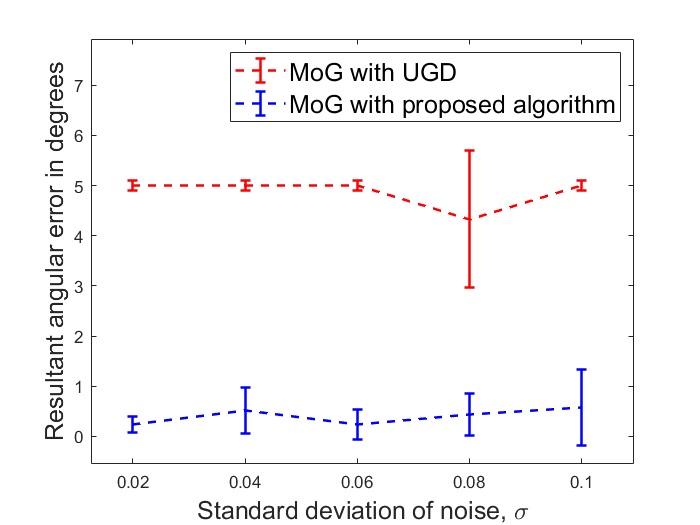}}
  \caption{Mean and standard deviation of resultant angular error vs varying noise when fibers are oriented at (a) $(\phi_1,\phi_2)=(10^\circ,50^\circ)$  and  (b) $(\phi_1,\phi_2,\phi_3)=(50^\circ,110^\circ,170^\circ)$. Polar angle is set to $90^\circ$ for all fibers.}
  \label{fig:8}
\end{figure}

\begin{figure}[!ht]
  \centering
  \subcaptionbox{\label{fig9:a}}{\includegraphics[width=2.1in]{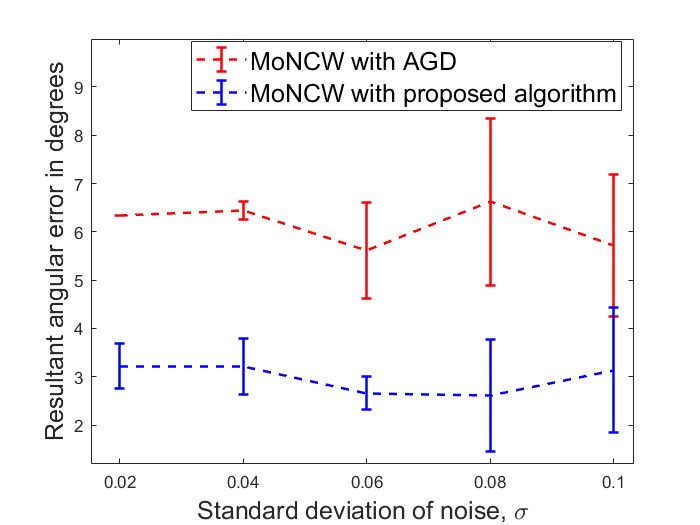}}
  \subcaptionbox{\label{fig9:b}}{\includegraphics[width=2.1in]{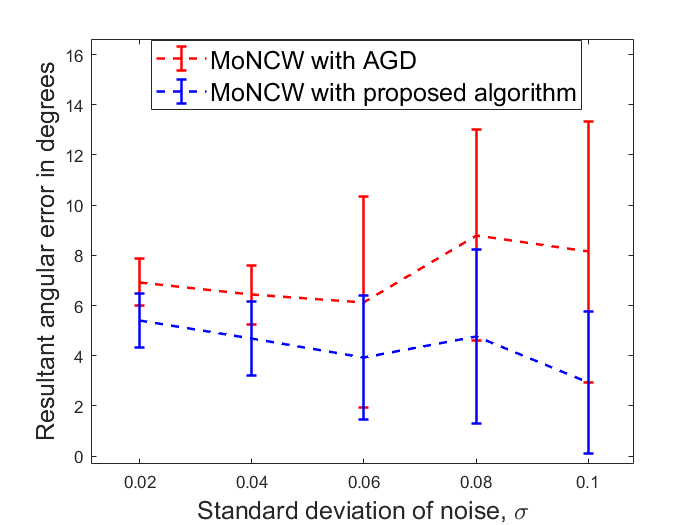}}
  \caption{Mean and standard deviation of resultant angular error vs varying noise when fibers are oriented at (a) $(\phi_1,\phi_2)=(10^\circ,70^\circ)$  and  (b) $(\phi_1,\phi_2,\phi_3)=(0^\circ,65^\circ,130^\circ)$. Polar angle is set to $90^\circ$ for all fibers.}
  \label{fig:9}
\end{figure}

In Fig. \ref{fig8:a} and \ref{fig9:a}, we plot mean and standard deviation of the resultant angular
errors for 2 crossing fibers with azimuthal angle $(\phi_1,\phi_2)=(10^\circ,50^\circ)$ and $(\phi_1,\phi_2)=(10^\circ,70^\circ)$  respectively and for 3 crossing fibers with azimuthal angle $(\phi_1,\phi_2,\phi_3)=(50^\circ,110^\circ,170^\circ)$ and $(\phi_1,\phi_2,\phi_3)=(0^\circ,65^\circ,130^\circ)$ in Fig. \ref{fig8:b} and Fig. \ref{fig9:b} respectively at varying noise level. The improvement in estimating the orientation of fibers using proposed model is significant at each noise level with both discussed models and gradient direction schemes. 

\subsection{\textbf{Analysis when crossing fibers do not reside in XY-plane}}

In the state-of-art models, the polar angles have always been set to $90^\circ$ indicating that the fibers are residing in the XY-plane. We have performed experiments with polar angles other than $90^\circ$. The analysis have been given in Table~\ref{table:3}. Here also we find huge reduction in angular error in all the cases using the proposed iterative approach.\\

\begin{table}[!ht] 
\centering
    \begin{tabular}{|c||c|c|c|c|}
   \hline
 $(\phi_1,\theta_1)-(\phi_2,\theta_2)$& MoNCW & MoNCW with & MoG with  & MoG with pro-\\
  (in degrees) & with AGD &  proposed algorithm & UGD &  posed algorithm\\
 
\hline
$(30^\circ,30^\circ)-(80^\circ,80^\circ)$& 12.42 &\textbf{ 4.74}&7.07&5.93\\

\hline

$(40^\circ,30^\circ)-(100^\circ,80^\circ)$ & 6.57 & \textbf{4.25}&4.43&4.57\\

\hline

$(45^\circ,25^\circ)-(105^\circ,95^\circ)$ & 6.25 & \textbf{0.96}&7.57&1\\

\hline

$(10^\circ,105^\circ)-(70^\circ,28^\circ)$ & 15 & \textbf{0.5}&3.93&\textbf{0.5}\\

\hline

$(10^\circ,48^\circ)-(20^\circ,100^\circ)$ & 8.39 & \textbf{5.17}&11.50&11.07\\

\hline

$(30^\circ,10^\circ)-(90^\circ,80^\circ)$ & 5.58 & \textbf{3.75}&3.79&5.35\\

\hline\hline
\end{tabular}

 \caption{Comparison of resultant angular error of $\phi$ and $\theta$ using
 different approaches with R-D noise, $\sigma=0.06$ for 2-crossing fibers not residing in XY-plane.}\label{table:3}
  
\end{table}

\subsection{\textbf{Experiment on Real Data}}

\begin{figure}[!ht]
  \centering
  
  \subcaptionbox{\label{fig10:a}}{\includegraphics[width=2.6in]{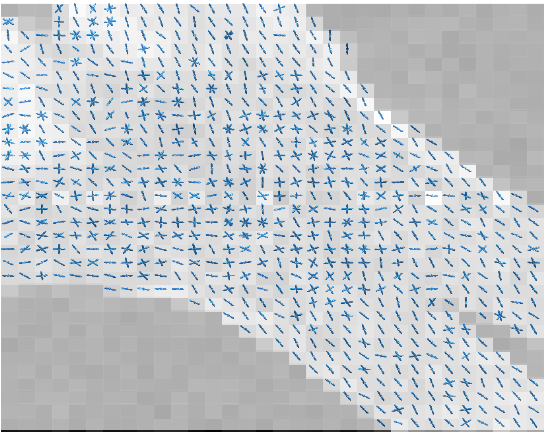}}
  \subcaptionbox{\label{fig10:b}}{\includegraphics[width=2.6in]{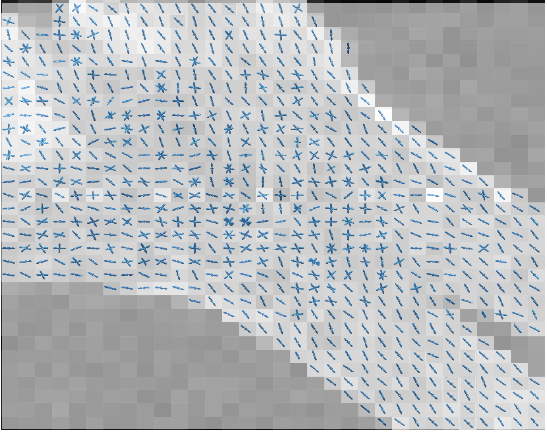}}
  \caption{Fiber orientational map of the optic chiasm of rat's brain using proposed algorithm when couples with (a)MoG model and (b) MoNCW model.}
  \label{fig:10}
\end{figure}

After synthetic data simulations, we conducted the real data experiment. We used the  optic chiasm of rat's brain's DW-MRI data set downloaded from \cite{barmpoutis2010tutorial} to corroborate the proposed model. Due to the well defined myelinated structure comprising of both parallel and intersecting optic nerve fibers, optic chiasm of rat's brain is an outstanding region to authenticate the proposed idea. The b-value for the data was $1250s/mm^2$. The other parameters of the experiment were set as $\Delta = 17.5ms$, $\delta = 1.5ms$ TE (Echo time) = $25ms$ and TR (Repetition time) = $1.17s$. DW-MRI data was collected for 46 gradient directions. An image with $b\approx 0s/mm^2$
was also collected. The fiber orientational maps for optic chiasm of rat's brain using the proposed algorithm coupled with MoG and MoNCW models is shown
in Fig. \ref{fig:10}

\section{Discussions}

The main emphasizes of this study is on MRI signals obtained after scanning and data acquisition. These signals have been processed to detect the orientation of fibers in brain.  We have developed an iterative algorithm for generating gradient directions that are used to detect the white matter fibers. The development emphasises on better quality of reconstructed images. AGD algorithm \cite{puri2021enhanced} is extended in the present study. Similar to AGD approach, proposed algorithm ensures that we work upon the directions that are in the neighbourhood of the expected orientation of the fibers. Doing this saves time and effort by avoiding the calculations involved in the superfluous directions in a particular voxel when uniform gradient directions are used. Unlike AGD algorithm, proposed work makes sure that at each iteration the point indicating the true orientation of fiber enters the newly generated grid of gradient directions. Increasing the number of iterations may increase the accuracy of the result a bit but in an expense of long computational time period for giving the final image. Hence, four iterations have been done in this study so as to get final orientational mapping image of brain in comparatively lesser time and with lesser computational complexity. \\

The robustness and consistency of the proposed algorithm can be proved with its performance at every noise level as shown in the figures and mean and standard deviation plots. Associating the iterative gradient direction approach with models like MoNCW  and MoG resulted in reduction of angular error when compared with UGD and AGD approaches with these models. This indicates the stability and flexibility of the algorithm.In Fig. \ref{fig3:a}, on account of high noise level, the AGD approach is showing 3 fibers per voxel in some simulations after reconstruction. It shows that AGD is hindered when highly noisy data is  available for reconstruction process. However, it can be observed that the proposed algorithm is sufficiently resistant to such noisy data. For the same reason, when 3-fibers with sufficiently large separation angles are considered in Fig. \ref{fig:6} for simulations then it is observed that the reconstruction of  $0^\circ$ fiber using AGD approach is disoriented in almost all the simulations unlike the proposed paradigm.\\

The related existing studies so far, have worked on the artificial simulations of fibers with fixed polar angles of $90^\circ$ and with equal weight (or volume) fraction. For manifesting the authenticity of the proposed algorithm in a more generalized manner, we performed experiments involving simulations of fibers not residing in the XY-plane. Huge reduction in angular error is seen when compared with the other state-of-art techniques. For this set-up, the iterative AGD approach when associated with MoNCW model shows maximum error of $5.17^\circ$ which is quite less then $12.42^\circ$ given by AGD approach. Similarly, comparing the statistics of angular error referred in Table~\ref{table:1} and \ref{table:2} we can predict that the reconstructed images using the proposed algorithm can depict the fibers directions more accurately. Reconstruction of fibers with unequal weight fraction in a voxel is done to see the performance of the proposed iterative algorithm. When unequal weight fractions are considered then it becomes cumbersome for AGD and UGD approach to detect the fiber with smaller weight value. In Fig. \ref{fig:4} and \ref{fig:5}, it can very well observed that the $0^\circ$  fiber orientation having lesser weight value is not reconstructed in many of the simulations. In almost all the cases the visualization shows that the proposed work gives better results when compared with other models and gradient direction schemes.\\

The performance of the proposed algorithm is more pronounced when it is coupled MoNCW model as shown in figures and tables. The reconstructed orientational maps of the optic chiasm of rat's brain using MoG and MoNCW model is shown in Fig. \ref{fig:10}. These maps present estimated fiber orientations in numerous regions of myelinated axons from 2-optic nerve bundles intersecting each other to reach their
respective contra-lateral optic tracts \cite{kumar2009multi}. As mentioned in \cite{kumar2009multi}, the reconstruction using proposed model is better as  single fibers as well as fiber crossings can be seen to be correctly rendered. Figure shows fibers of cingulum and corpus callosum intersecting
each other at the middle region where the 2-fiber bundles seem to be crossing each other.\\

In conclusion, iterative gradient directions approach is developed to fill the gaps of existing models and gradient direction schemes. In this paper, we have developed a novel iterative technique for generating gradient directions in order to estimate the passage of white matter fibers in the brain or in knowing how the brain is wired. The results procured using proposed algorithm has been compared with the results of AGD and UGD approaches. In almost all the cases, huge reduction in angular error is found. The proposed algorithm also performed exceedingly well when synthetic simulation experiments for the fibers not residing in the XY-plane is performed. This model is promising for distinguishing crossing fibers at every noise level. The proposed algorithm for choosing gradient directions can be used in other models also, as the performance in both the models is superior to state-of-
art techniques. 

\section*{Statements of ethical approval and competing interests}

Present study does not require ethical approval as open source data set is used. The authors declare that they have no conflict of interest.

\section*{Acknowledgements}

One of the authors, Ashishi Puri, is grateful to the Ministry of Human Resource Development, India and the Indian Institute of Technology, Roorkee for financial support, to carry out this work. The grant number is MHR01-23-200-4028. This work is also supported by the project grant no. DST/INT/CZECH/P-10/2019 under Indo-Czech Bilateral Research Program.

\bibliographystyle{elsarticle-num}
\bibliography{bib}

\begin{thebibliography}{10}
\expandafter\ifx\csname url\endcsname\relax
  \def\url#1{\texttt{#1}}\fi
\expandafter\ifx\csname urlprefix\endcsname\relax\def\urlprefix{URL }\fi
\expandafter\ifx\csname href\endcsname\relax
  \def\href#1#2{#2} \def\path#1{#1}\fi

\bibitem{basser1994mr}
P.~J. Basser, J.~Mattiello, D.~LeBihan, Mr diffusion tensor spectroscopy and
  imaging, Biophysical journal 66~(1) (1994) 259--267.

\bibitem{mori2006principles}
S.~Mori, J.~Zhang, Principles of diffusion tensor imaging and its applications
  to basic neuroscience research, Neuron 51~(5) (2006) 527--539.

\bibitem{mori1999diffusion}
S.~Mori, P.~B. Barker, Diffusion magnetic resonance imaging: its principle and
  applications, The Anatomical Record: An Official Publication of the American
  Association of Anatomists 257~(3) (1999) 102--109.

\bibitem{luypaert2001diffusion}
R.~Luypaert, S.~Boujraf, S.~Sourbron, M.~Osteaux, Diffusion and perfusion mri:
  basic physics, European journal of radiology 38~(1) (2001) 19--27.

\bibitem{jones2011diffusion}
D.~K. Jones, A.~Leemans, Diffusion tensor imaging, in: Magnetic resonance
  neuroimaging, Springer, 2011, pp. 127--144.

\bibitem{soares2013hitchhiker}
J.~Soares, P.~Marques, V.~Alves, N.~Sousa, A hitchhiker's guide to diffusion
  tensor imaging, Frontiers in neuroscience 7 (2013) 31.

\bibitem{alexander2007diffusion}
A.~L. Alexander, J.~E. Lee, M.~Lazar, A.~S. Field, Diffusion tensor imaging of
  the brain, Neurotherapeutics 4~(3) (2007) 316--329.

\bibitem{westin2014measurement}
C.~F. Westin, F.~Szczepankiewicz, O.~Pasternak, E.~{\"O}zarslan, D.~Topgaard,
  H.~Knutsson, M.~Nilsson, Measurement tensors in diffusion mri: generalizing
  the concept of diffusion encoding, in: International conference on medical
  image computing and computer-assisted intervention, Springer, 2014, pp.
  209--216.

\bibitem{tuch2002high}
D.~S. Tuch, T.~G. Reese, M.~R. Wiegell, N.~Makris, J.~W. Belliveau, V.~J.
  Wedeen, High angular resolution diffusion imaging reveals intravoxel white
  matter fiber heterogeneity, Magnetic Resonance in Medicine: An Official
  Journal of the International Society for Magnetic Resonance in Medicine
  48~(4) (2002) 577--582.

\bibitem{jian2007multi}
B.~Jian, B.~C. Vemuri, Multi-fiber reconstruction from diffusion mri using
  mixture of wisharts and sparse deconvolution, in: Biennial International
  Conference on Information Processing in Medical Imaging, Springer, 2007, pp.
  384--395.

\bibitem{jian2007novel}
B.~Jian, B.~C. Vemuri, E.~{\"O}zarslan, P.~R. Carney, T.~H. Mareci, A novel
  tensor distribution model for the diffusion-weighted mr signal, NeuroImage
  37~(1) (2007) 164--176.

\bibitem{shakya2017multi}
S.~Shakya, N.~Batool, E.~{\"O}zarslan, H.~Knutsson, Multi-fiber reconstruction
  using probabilistic mixture models for diffusion mri examinations of the
  brain, in: Modeling, Analysis, and Visualization of Anisotropy, Springer,
  2017, pp. 283--308.

\bibitem{shakya2017multitrac}
S.~Shakya, X.~Gu, N.~Batool, E.~{\"O}zarslan, H.~Knutsson, Multi-fiber
  estimation and tractography for diffusion mri using mixture of non-central
  wishart distributions, in: Eurographics Workshop on Visual Computing for
  Biology and Medicine, September 7-8, 2017, Bremen, Germany, The Eurographics
  Association, 2017, pp. 1--5.

\bibitem{schilling2020brain}
K.~G. Schilling, L.~Petit, F.~Rheault, S.~Remedios, C.~Pierpaoli, A.~W.
  Anderson, B.~A. Landman, M.~Descoteaux, Brain connections derived from
  diffusion mri tractography can be highly anatomically accurate—if we know
  where white matter pathways start, where they end, and where they do not go,
  Brain Structure and Function 225~(8) (2020) 2387--2402.

\bibitem{schucht2020visualization}
P.~Schucht, H.~R. Lee, H.~M. Mezouar, E.~Hewer, A.~Raabe, M.~Murek, I.~Zubak,
  J.~Goldberg, E.~K{\"o}vari, A.~Pierangelo, et~al., Visualization of white
  matter fiber tracts of brain tissue sections with wide-field imaging mueller
  polarimetry, IEEE transactions on medical imaging 39~(12) (2020) 4376--4382.

\bibitem{puri2021enhanced}
A.~Puri, S.~Shakya, S.~Kumar, An enhanced multi-fiber reconstruction technique
  using adaptive gradient directions coupled with moncw model in diffusion mri,
  Journal of Magnetic Resonance (2021) 106931.

\bibitem{stejskal1965spin}
E.~O. Stejskal, J.~E. Tanner, Spin diffusion measurements: spin echoes in the
  presence of a time-dependent field gradient, The journal of chemical physics
  42~(1) (1965) 288--292.

\bibitem{james1955non}
A.~T. James, The non-central wishart distribution, Proceedings of the Royal
  Society of London. Series A. Mathematical and Physical Sciences 229~(1178)
  (1955) 364--366.

\bibitem{li2003noncentral}
K.~Li, Z.~Geng, The noncentral wishart distribution and related distributions,
  Communications in Statistics-Theory and Methods 32~(1) (2003) 33--45.

\bibitem{letac2004tutorial}
G.~Letac, H.~Massam, A tutorial on non central wishart distributions, Technical
  Paper, Toulouse University (2004).

\bibitem{pham2015trace}
T.~Pham-Gia, D.~N. Thanh, D.~T. Phong, et~al., Trace of the wishart matrix and
  applications, Open Journal of Statistics 5~(03) (2015) 173.

\bibitem{mayerhofer2013existence}
E.~Mayerhofer, On the existence of non-central wishart distributions, Journal
  of Multivariate Analysis 114 (2013) 448--456.

\bibitem{lawson1995solving}
C.~L. Lawson, R.~J. Hanson, Solving least squares problems, SIAM, 1995.

\bibitem{gindikin1975invariant}
S.~G. Gindikin, Invariant generalized functions in homogeneous domains,
  Functional analysis and its applications 9~(1) (1975) 50--52.

\bibitem{n1988davidson}
D.~N.~Shanbhag, The davidson-kendall problem and related results on the
  structure of the wishart distribution, Australian Journal of Statistics
  30~(1) (1988) 272--280.

\bibitem{peddada1991proof}
S.~D. Peddada, D.~S.~P. Richards, et~al., Proof of a conjecture of ml eaton on
  the characteristic function of the wishart distribution, The Annals of
  Probability 19~(2) (1991) 868--874.

\bibitem{barmpoutis2009adaptive}
A.~Barmpoutis, B.~Jian, B.~C. Vemuri, Adaptive kernels for multi-fiber
  reconstruction, in: International Conference on Information Processing in
  Medical Imaging, Springer, 2009, pp. 338--349.

\bibitem{barmpoutis2010tutorial}
A.~Barmpoutis, Tutorial on diffusion tensor mri using matlab, Electronic
  Edition, University of Florida (2010).

\bibitem{kumar2009multi}
R.~Kumar, B.~C. Vemuri, F.~Wang, T.~Syeda-Mahmood, P.~R. Carney, T.~H. Mareci,
  Multi-fiber reconstruction from dw-mri using a continuous mixture of
  hyperspherical von mises-fisher distributions, in: International Conference
  on Information Processing in Medical Imaging, Springer, 2009, pp. 139--150.

\end{thebibliography}
\end{document}